\documentclass[apj]{aastex6}

\shorttitle{Electron and proton modulation during solar minimum 2006-2009}
\shortauthors{Di Felice et al.}

\newcommand{\el}{\mbox{${\rm e^{-}}\; $}}
\newcommand{\pos}{\mbox{${\rm e^{+}}\;$}}

\newcommand{\ep}{\mbox{${\rm e^{-}}/{\rm p}\;$}}

\begin{document}

\title{New evidence for charge-sign dependent modulation during the solar minimum of 2006 to 2009} 

\author{
  V. Di Felice$^{1,2}$, R.~Munini$^{3}$, E. E. Vos$^{4}$, and M. S. Potgieter$^{4}$
}

\affil{$^{1}$INFN, Sezione di Roma ``Tor Vergata'', I-00133 Rome, Italy}
\affil{$^{2}$Agenzia Spaziale Italiana (ASI) Science Data Center, Via del Politecnico snc, I-00133 Rome, Italy}
\affil{$^{3}$INFN, Sezione di Trieste, I-34149 Trieste, Italy}
\affil{$^{4}$Centre for Space Research, North-West University, 2520 Potchefstroom, South Africa}

\begin{abstract}

The PAMELA space experiment, in orbit since 2006, has measured cosmic rays through the most recent $A<0$ solar minimum activity period.
During this entire time, galactic electrons and protons have been detected down to $70$~MV and $400$~MV, respectively, and their differential intensity variation in time has been monitored with unprecedented accuracy. These observations are used to show how differently electrons and protons responded to the quiet modulation conditions that prevailed from 2006 to 2009.
It is well known that particle drifts, as one of four major mechanisms for the solar modulation of cosmic rays, cause charge-sign dependent solar modulation. 
Solar minimum activity periods provide optimal conditions to study these drift effects. The observed behaviour is compared to the solutions of a three-dimensional model for cosmic rays in the heliosphere, including drifts. The numerical results confirm that the difference in the evolution of electron and proton spectra during the last prolonged solar minimum
is attributed to a large extent to particle drifts.
We therefore present new evidence of charge-sign dependent solar modulation, with perspective on its peculiarities for the observed period from 2006 to 2009.
\end{abstract}

\keywords{Cosmic rays; solar wind; Sun; heliosphere; drifts}

\section{Introduction}
\label{Sec:intro}
When entering the heliosphere, charged particles constituting cosmic rays (CRs) of galactic origin interact with the turbulent solar
wind and its embedded heliospheric magnetic field (HMF). They undergo convection, diffusion and adiabatic energy losses while 
traversing the expanding solar wind.  They also sense the gradients and curvatures in the global HMF, and the effect of the heliospheric current sheet (HCS), causing them to drift according to the polarity of the HMF. The resulting solar modulation significantly modifies the local interstellar cosmic ray spectra in intensity and shape, a process that depends on the type of particles, their energy, sign of charge, and solar activity, both in terms of space, where in the heliosphere they are observed, and in time, when during solar cycles they are measured. 
CRs with energies up to tens of GeV are affected, progressively more with decreasing energies so that below a few GeV their solar modulation becomes significantly large; 
see~\citet{Strauss_Potgieter_2014a}. 

At energies above 10 GeV, evidence for CR modulation has been provided for decades by ground level CR detectors, called neutron monitors.
These observations show a clear anti-correlation between the CR intensity and solar activity for the whole 11-year solar cycle, from its minimum,
when the Sun is quiet and the CR intensity is at its largest, to its maximum when the CRs reach their minimum intensity. A $22$-year
periodicity is also evident in these observations caused by the polarity reversal of the HMF, which takes place around extreme maximum solar activity, every $\sim~11$ years. 
This 22-year cycle in cosmic rays, which is not at all evident in e.g. sunspot numbers as a proxy for solar activity, is a manifestation of the effects of particle drifts.
In fact, during an $A>0$ polarity cycle of the HMF, when magnetic field lines are pointing outward in the northern heliohemisphere, positively charged particles
drift into the inner heliosphere mainly through the polar regions of the heliosphere and then outwards mainly along the wavy HCS. This drift pattern reverses when the HMF changes its polarity so that during $A<0$ cycles positively charged particles reach the Earth mainly through the equatorial regions, directly encountering the wavy HCS in the process. 
Negatively charged particles then drift inwards mainly through the polar regions. This means that protons and electrons sense different regions of the heliosphere during the same polarity period while travelling through the heliosphere to the Earth, providing a clear signature for charge-sign dependent modulation; for an elaborate discussion of these effects, see the reviews by~\citet{Potgieter_2013,Potgieter_2014}.

Since CRs are responsive to heliospheric modulation conditions, they can be used very effectively to provide fundamental information for our understanding
 of the features of charged particle transport and also the details of solar modulation, because the interplay between the mentioned major modulation 
mechanisms changes with solar activity. Moreover, precise CR observations are crucially important when addressing fundamental questions in astrophysics, 
e.g. particle and anti-particle CR data can be utilised to search for hints of new physics like dark matter: 
beside the high energy antiparticle data~\citep{adriani2013_PRL111, Adriani_2009_nature} also their low energy spectra are interesting since they can be used~\citep[e.g.][]{Hooper_2015, Cerde_o_2012, Bottino_2012} to constrain models with light ($\sim 10$~GeV) dark matter candidates put forward to interpret measurements by direct-detection experiments~\citep[e.g.][]{Aalseth_2011, Belli_2011}.  These data can be fully exploited with a precise understanding of the expected background due to the production of antiparticles by CR interaction with the interstellar matter and their transport to the Earth.
Consequently, this necessitate that the charge-sign dependent solar modulation and how it changes with time are accurately interpreted.
The importance of solar modulation is clearly demonstrated by the large variability of CR measurements at energies below a few tens of GeV performed inside the heliosphere by several space and balloon missions over the years, such as the Voyagers~\citep{Stoneetal2013}, Ulysses~\citep{Simpson_1992, Heber_2009} and balloon flights~\citep[see e.g.][and references therein]{Seo_2012}. 

Since the late 60' observations made at Earth have been interpreted using modelling of the solar modulation with the force field approximation~\citep{Gleeson_1968}.
However this approach is severely limited in exploring and subsequently explaining the full range of processes responsible for the solar modulation of CRs. 
This approach is valid for only one spatial dimension so that it cannot account for any process such as diffusion perpendicular to the HMF and for particle drifts which essentially requires a full 3D approach. The importance of drifts was already illustrated and emphasized in the late 1970's, see ~\citet{Jokipii_1977}, and early 1980's with elaborate numerical illustrations by~\citet{JokipiiKota1985} and~\citet{Potgieter_Moraal_1985}, who also applied their model to charge-sign dependent observations; see the recent review by~\citet{Potgieter_2014}. 
Experimental evidence of charge-sign dependent solar modulation has gradually build up over the years. Measurements of proton, helium and the sum of electron and positron intensities, performed in the inner heliosphere by balloon-based experiments and spacecraft, during cycles of opposite polarity,
provided evidence of this effect. Moreover, positron fraction measurements~\citep{Clem_2009} clearly show how relevant charge-sign dependent modulation is below a few GeV. Alongside experimental findings, significant progresses have been made in drift theory, supported by increasingly complex and accurate modelling. These models can describe a full 22-year solar cycle~\citep{Le_Roux_Potgieter_1995} and even predict how drift should change with the solar activity cycle ~\citep{Ferreira_2004}. In addition, drift effects on radial and latitudinal gradients were illustrated \citep{Potgieter_1989_JGR, VosPotgieter2016}. Comprehensive modelling of drift effects was done on the solar modulation of protons, electrons, positrons and even anti-protons \citep{Webber_Potgieter_1989,Potgieter_Langner_2004, Langner_Potgieter_2004}.

In order to improve our understanding of the details of the interplay among the several solar modulation  mechanisms, in particular the importance of drifts during solar minimum epochs and over the entire solar cycle, precise and simultaneous measurements are required of CRs with opposite charge-signs. Nowadays, such precise measurements are being performed in space by the magnetic spectrometers PAMELA~\citep{adriani2014pamela}, a satellite-borne experiment, and by AMS-02~\citep{Battiston_2010}, on board of the International Space Station. 
Fortunately, the PAMELA experiment has obtained data during the entire peculiar and extraordinary prolonged $A<0$ solar minimum of cycle 23/24, 
that lasted until the end of $2009$~\citep[][and references therein]{Potgieter_2013}. This mission has been detecting CR spectra for protons, electrons and positrons from a few tens of MeV to several hundreds of GeV. A report of how the $e^{+}/e^{-}$ ratios between 0.5 and 5.0 GeV changed from 2006 to 2015 has recently been made~\citep{Adriani_2016_elpos_mod}. Here, we show the temporal evolution of electron and proton spectra during the last solar minimum, in conjunction with comprehensive modelling, exploring and utilizing the extended energy range and improved precision.
These differences will be displayed respectively as proton and electron time profiles, from 2006 to 2009, together with the \ep ratio as a function of rigidity for the mentioned period. 
Evidently, with the observation of precise spectra for protons, electrons,  positrons and even anti-protons, on an almost continuous time-scale, charge-sign dependent modulation, and other important 3D effects, should no longer be ignored when interpreting these data, so that surpassing the force-field model has become necessary; see also ~\citet{Maccione_2013, Cholis_2016}.

\section{The PAMELA electron and proton differential fluxes}
\label{Sec:data}

The PAMELA spectrometer was conceived and built to study the antimatter component of CRs over a wide energy range, from tens of MeV up to
hundreds of GeV, significantly improving collected statistics and precision with respect to previous experiments. The magnetic spectrometer, 
composed of a permanent magnet and a tracking system of six planes of double-sided silicon sensors, reconstructs the 
particle trajectory and determines its curvature, distinguishing oppositely charged particles; it also provides a measure of the particle rigidity 
 $R$ =$ pc/Ze$ ($p$ and $Ze$ being the particle momentum and charge, respectively, and $c$ the speed of light) and of ionization energy losses.
The particle velocity is obtained combining the time of passage information given by the Time-of-Flight system with the track lenght. 
 With these information the spectrometer
can distinguish between down-going particles and up-going splash-albedo particles and separate negatively from positively charged
particles.   Hadron-lepton separation is achieved thanks to a sampling imaging calorimeter. A shower tail catcher and a neutron detector 
beneath help in the discrimination, while an anticoincidence system is used to reject spurious events. More details on the instrument can be found in~\citet{Picozza_2007}.

Such apparatus is optimized for the study of charge one particles and to reach a high level of electron-proton discrimination.
Thus, aside from antimatter studies, PAMELA also performed precise measurements of the matter component of cosmic rays~\citep{adriani2014pamela}. 
In particular, galactic electron and proton differential fluxes have
 been measured up to $\sim600$~GV and $\sim1$~TV respectively~\citep{Adriani_2011_el, Adriani_2011_science}, and 
down to energies as low as few tens of MeV. 
Since its launch in June 2006, the instrument is following a quasi-polar orbit at an inclination of  $70^\circ$,  sampling low geomagnetic cutoff regions
at high latitude. Thus, the lowest detectable rigidity limit is not due to geomagnetic cutoff effects, but to the high curvature of
low energy particles in the instrument magnetic field, which causes them not to trigger.

The long duration flight allowed to collect data 
during the entire 2006-2009 $A<0$ solar minimum, continuously monitoring    
the flux of charged particles of galactic origin.
\citet{Adriani_2013_pmodul} reported the proton flux time-dependent measurements, performed by PAMELA 
down to $400$~MV, which are relevant for solar modulation studies and are used in this work. 
The large statistics collected allowed to measure the average proton flux over each Carrington rotation~\citep{Carrington_1863},
from number 2045 to 2092 (hereafter referred as Carrington flux). 
We also use galactic electron flux data 
measured by PAMELA during the same period, as reported by ~\citet{Adriani_2015_elmodul}. 
Due to the lower statistics the \el fluxes were measured 
over a six-month time basis. PAMELA data are obtained through the Cosmic Ray Data Base of the ASI Science Data Center~\citep{CRDB}.

In order to compare the spectral evolution of protons and electrons, 
proton data have been combined on a semestral time basis, conforming to the temporal 
division of the electron measurements. 
For each semester a galactic proton differential flux, $\varphi$, has been obtained by performing a weighted average of the 
relevant Carrington fluxes, $\varphi_i$:

\begin{equation}
\varphi  = \frac{\sum_{i=1}^{N}  \varphi_i w_i} {  \sum_{i=1}^{N} w_i}
\end{equation}
where $i$ is the index running on the number $N$ of Carrington rotations within the considered semester, 
and the weight $w_i$ is the number of active days of the instrument for the $i-th$ Carrington rotation.
This procedure resulted in $7$ differential proton fluxes as a function of kinetic energy for each semester from July 2006 to December 2009.

\begin{figure}[tb] 

  \begin{minipage}{1.\textwidth}
    \centering
    \plotone{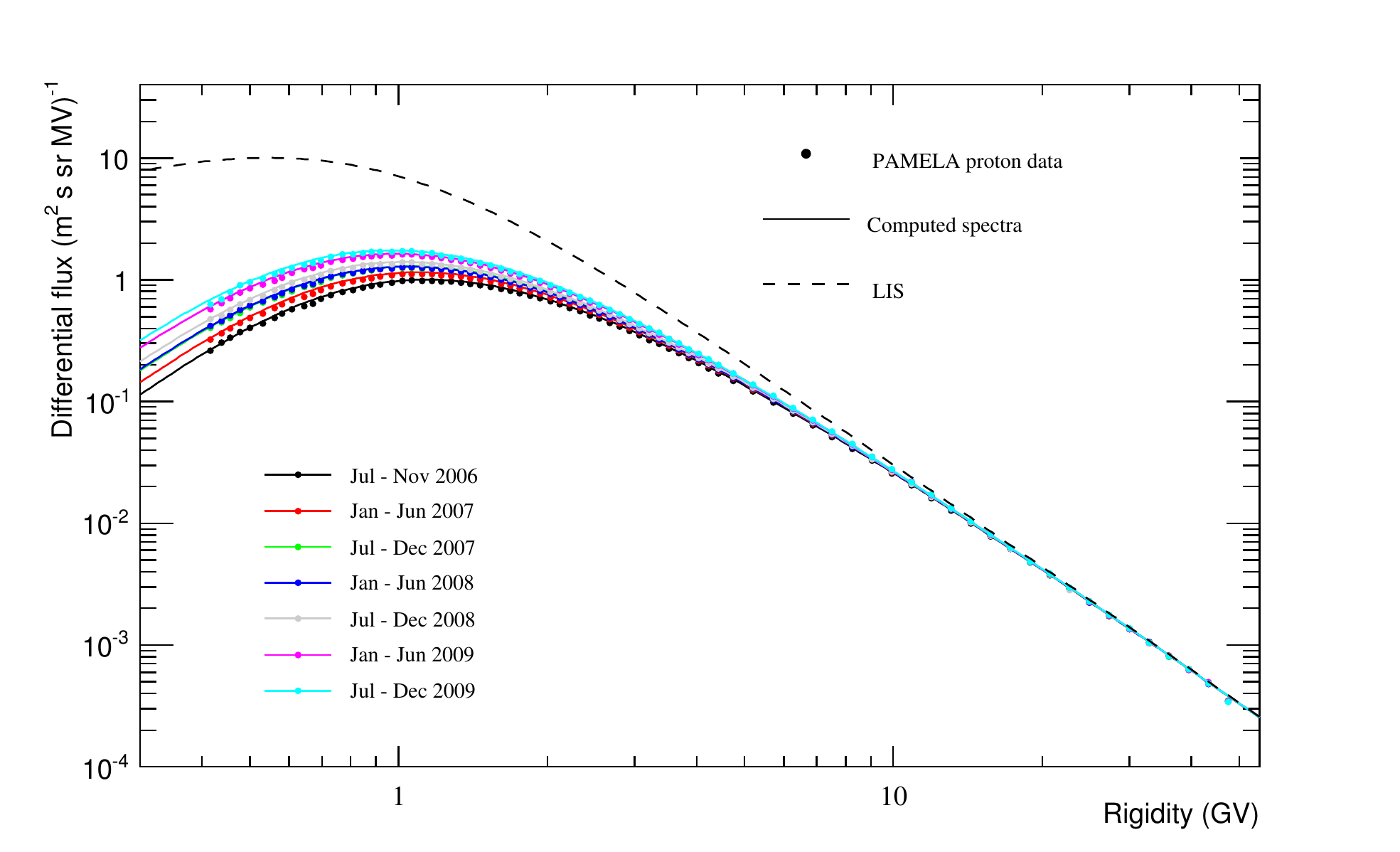}
    \vspace{1em}
  \end{minipage}

  \begin{minipage}{1.\textwidth}
    \centering 
    \plotone{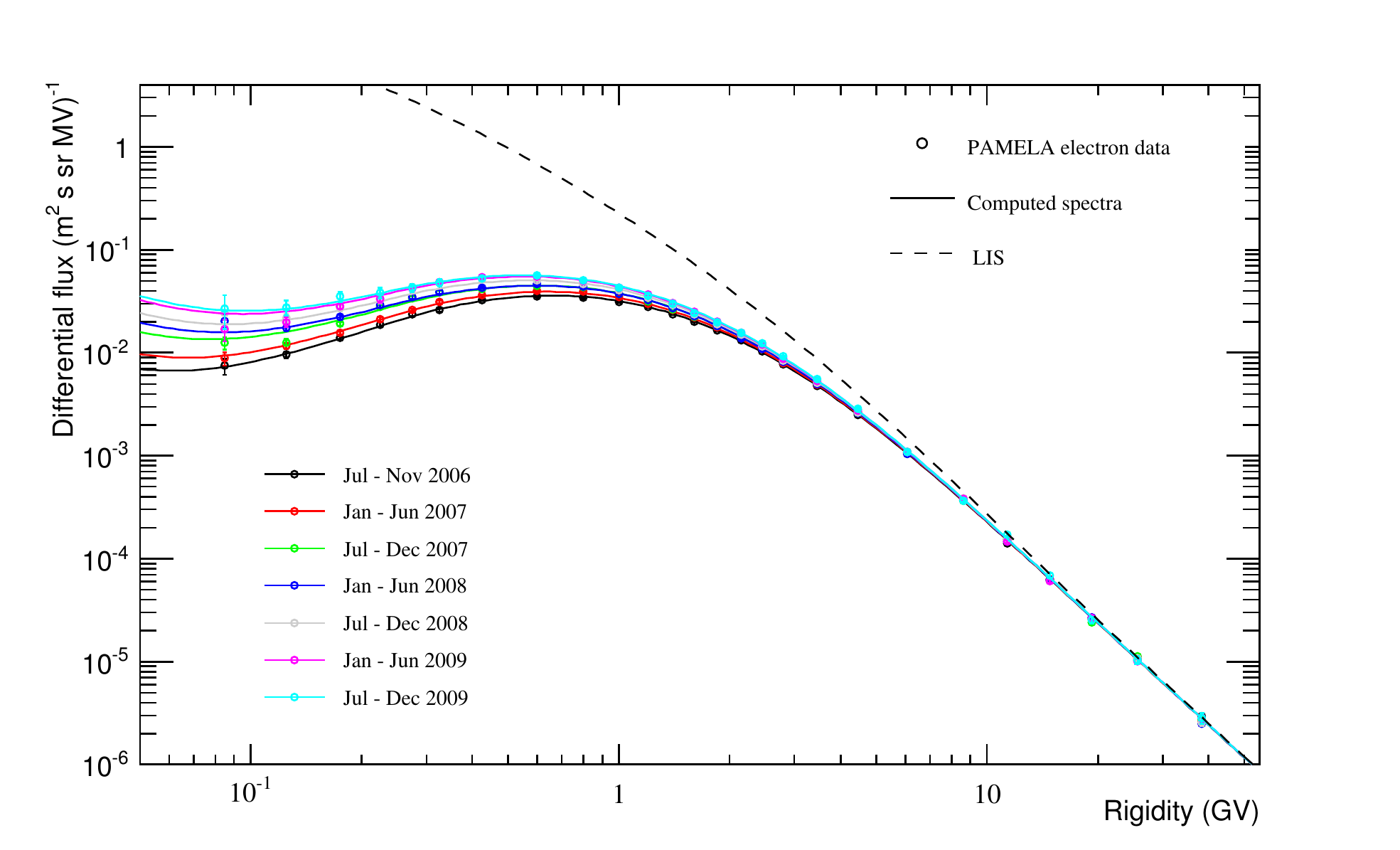}
  \end{minipage}  
  
\caption{Six-month averaged proton (upper panel) and electron (lower panel) spectra as measured by PAMELA from mid-2006 to the end of 2009~\citep{Adriani_2013_pmodul,Adriani_2015_elmodul} are shown as a function of the rigidity. Observations are overlaid with the corresponding computed spectra (continuous lines) for an A$<$0 cycle magnetic field polarity~\citep{Potgieter_2014_pmodul, VosPotgieter_2015, Potgieter_2015_elmodul}. Colour coding shown in the legend indicates the seven selected time slots for both experimental and computed spectra. The proton and electron LISs (dashed lines) used for the computation are also shown.}
\label{fig:H_el_fluxes}
\end{figure}

For a quantitative understanding of the different temporal development of the electron and proton spectrum 
it is necessary to compare their recovery to solar minimum at the same rigidities.
Particle spectra available as fluxes in kinetic energy, $\varphi(E)$ in units of $(\text{m}^2 \text{s sr MeV})^{-1}$, 
have been converted to fluxes in rigidity, $\varphi(R)$ in units of
$(\text{m}^2 \text{s sr MV})^{-1}$, according to the following:

 \begin{equation}
\varphi(R)  = \varphi(E) \frac{R} { \sqrt{R^2 + m^2}}
\label{eq:rig_en_conversion}
\end{equation}
with $m$ the particle mass. The application of Eq.~\ref{eq:rig_en_conversion} in the PAMELA measurement energy range results in a significant modification of the 
proton spectral shape below few tens of GeV, while the effect on the electron flux is negligible due to their lower mass. 
The resulting proton and electron spectra as a function of rigidity are shown
 in Figure~\ref{fig:H_el_fluxes}. 
Error bars represent the statistical errors. Because of statistical requirements,
 electrons and protons were measured in different rigidity intervals in the whole range of interest, and this will be detailed in Section~\ref{Sec:discussion} when comparing the temporal evolution of particle intensities in the whole rigidity range.
Solid lines in these figures are the solutions of a comprehensive numerical modulation model, reproducing the particle differential spectra for 
each period as indicated.
For both electrons and protons the newly determined local interstellar spectra, see also ~\citet{Potgieter_2014_Brazil}, used as unmodulated input spectra for the model are also shown in the two figures. 

The different behaviour of electrons in comparison with protons at low rigidities were discussed in detail
 by \citet{Potgieter_2014_pmodul,Potgieter_2015_elmodul} and is not repeated here. 
In the rest of our work, we focus on rigidities above $0.4$~GV, which is the lowest rigidity available for the proton spectra. 
Relevant aspects of the theory and modelling are concisely described in Sec.~\ref{Sec:model} 
of this paper. For both CR species, the low-energy part of the spectra varies significantly with time, responding to changes in solar activity during the observed $23/24$  cycle solar minimum period. As expected, the lowest energies are the most responsive. 

\section{Modulation theory and numerical modelling}
\label{Sec:model}
A comprehensive 3D numerical model was implemented to study the changing proton and electron fluxes measured by PAMELA during the $23/24$ solar minimum, following similar studies by~\citet{Potgieter_2014_pmodul} and~\citet{Potgieter_2015_elmodul}. 
The model used in this study is based on the \citet{Parker1} transport equation, which describes the transport and modulation of CRs in the heliosphere. The advantage of this model over traditional force-field based models is that all of the important modulation processes are explicitly accounted for, namely diffusion in all dimensions, convection, adiabatic energy losses, and drifts, which allows for an in-depth study of heliospheric modulation.  See \citet{Potgieter2013} for a detailed overview of heliospheric modulation and the various modulation processes.
\par
Using the above model, energy spectra were computed that reproduce PAMELA proton and electron measurements, as shown in Figure~\ref{fig:H_el_fluxes}.
 The heliopause (HP) for the simulated heliosphere was taken at $122\,$AU, with a termination shock (TS) position that varies between $88\,$AU in 2006, and $80\,$AU in 2009.
Newly determined local interstellar spectra (LIS) are specified for protons and electrons at the HP, which serve as input spectra at this modulation boundary. Both the electron and proton LISs were constructed to match Voyager $1$ measurements below $100\,$MeV, taken from beyond the HP in August 2012, and PAMELA measurements above $\sim30\,$GeV, where measurements are expected to become less affected by solar modulation~\citep[e.g.][]{StraussPotgieter2014b}.  This approach leads to reliable estimates for both the proton and electron LIS~\citep[see also][]{Potgieter_2015_elmodul, VosPotgieter_2015}.
\par
\begin{figure}
\epsscale{.85}
\plotone{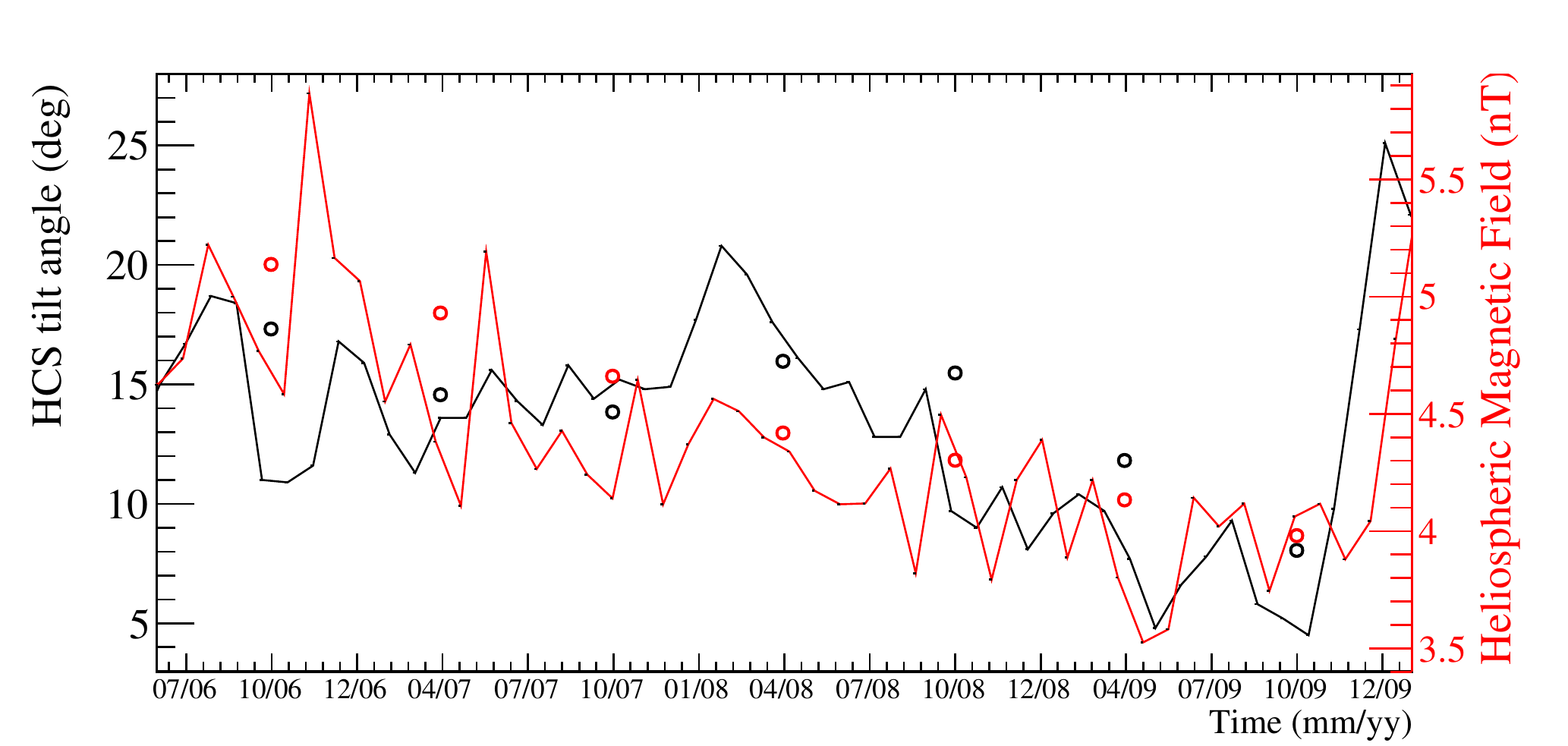}
\caption{Temporal evolution of the HCS tilt angle (black line) and the HMF (red line) from mid-2006 to the end of 2009.
 Data are taken from~\citet{WSO} and \citet{HMF}, respectively.  The open  points represent the average values for the preceding 12 months,
 used as input for the modelling to compute the seven averaged electron and proton spectra.  See also~\citet{VosPotgieter_2015}.}
\label{fig:tilt_B}
\end{figure}

The HCS tilt angle, also seen as a proxy for solar activity, and the magnitude of the HMF at the Earth changed during the years leading up to 2009.  Both are important entities for describing drift modulation. When reproducing the PAMELA proton and electron spectra these changes were accounted for by setting up realistic modulation conditions in the model that coincide with the semesterly averaged spectra from PAMELA.  Averages for the HCS tilt angle and the HMF were calculated in order to obtain representative values for these parameters that are indicative of preceding modulation conditions. 
Both the HCS tilt angle and the HMF used in the model are shown in Figure~\ref{fig:tilt_B}.
The HMF became more ordered over the years leading up to 2009, which translates to a reduction in the amount of turbulence in the heliosphere and to subsequent increases in the particle mean free paths (MFPs).  These increases, along with gradient, curvature and current sheet drifts, are expected to be responsible for the proton and electron intensity increases observed by PAMELA from 2006 to 2009.  In this study the HMF is described according to \citet{SmithBieber1991}.
\par
The numerical solutions (lines) shown in Figure~\ref{fig:H_el_fluxes} 
 were obtained using a diffusion approach that approximates quasi-linear theory (QLT), while still adhering to constraints from more advanced turbulence studies~\citep[e.g.][]{Potgieter2000, Shalchi2009}. 
The equation for diffusion parallel to the average background HMF is given by
\begin{equation}
  \kappa_\parallel = \kappa_{\parallel{0}}\beta\,{F}(r,\theta,\phi)\,G(R),
\end{equation}
with $\kappa_{\parallel{0}}$ a constant in units of cm$^2$s$^{-1}$ and $\beta$ the ratio of particle speed to the speed of light.  $F(r,\theta,\phi)$ is a function that provides a $B^{-1}$ spatial dependence for the diffusion coefficients (DCs), with $B$ the HMF magnitude at a given position in the heliosphere, and $r$, $\theta$ and $\phi$ the radial distance, polar angle and azimuthal angle, respectively.  $G(R)$ is a function that takes care of the rigidity dependence, which consists of two combined power-laws.  For diffusion perpendicular to the HMF lines, distinction is made between the radial ($\kappa_{\perp{r}}$) and polar ($\kappa_{\perp\theta}$) directions, where the former and latter are scaled to $2\,$\% and $1\,$\% of $\kappa_\parallel$, respectively.  See \citet{VosPotgieter_2015} for detailed discussions on this diffusion approach.
\par
Of particular importance to this study is particle drifts, which are caused by the presence of gradients and curvatures in the HMF, as well as by the sudden HMF polarity change across the HCS.  In the weak-scattering limit, which translates to the largest possible drift effects from drift theory, the expression for the average guiding center drift velocity is given by
\begin{equation}
  \langle\mathbf{v_D}\rangle = \nabla\times\kappa_D\mathbf{e_B},
\label{eq:eq4}
\end{equation}
with $\mathbf{e_B}=\mathbf{B}/B$ a unit vector directed along the HMF vector $\mathbf{B}$, and $\kappa_D$ a generalized drift coefficient that is related to the drift scale ($\lambda_D$) and the particle speed ($v$) by $\lambda_D=3\kappa_D/v$.  The weak-scattering approach was found to be too simple for the purpose of this study, so that a modification had to be applied to $\kappa_D$ which takes into account the effects of scattering on drifts, leading to smaller drift scales at lower rigidities \citep[e.g.][]{Burgeretal2000}.  The drift coefficient is therefore given by
\begin{equation}
  \kappa_D = \kappa_{D0}\,\frac{\beta{R}}{3B}\,\frac{\left(\frac{R}{R_{D0}}\right)^2}{1+\left(\frac{R}{R_{D0}}\right)^2},
\label{eq:eq5}
\end{equation}
with $\kappa_{D0}$ a constant that determines the amount of drifts (taken here as $1.0$ for $100\,$\% drift effects), and $R_{D0}=0.55\,$GV a constant in GV that determines the rigidity below which $\kappa_D$ is modified.  Deviating from the weak-scattering drift approach prevents the overestimation of drift effects, even under perfect solar minimum conditions, as was present during 2009. 
This approach of reducing drifts at lower rigidities was originally introduced to explain the very small latitudinal CR gradients
observed by Ulysses \citep{Heber_2002, HeberPotgieter2006, Heber_2008}. In the work presented here, this particular reduction in terms of rigidity above $0.4$~GV is less important.
\par
\begin{figure}[!tbp]
  \centering
  \includegraphics[width=0.49\textwidth]{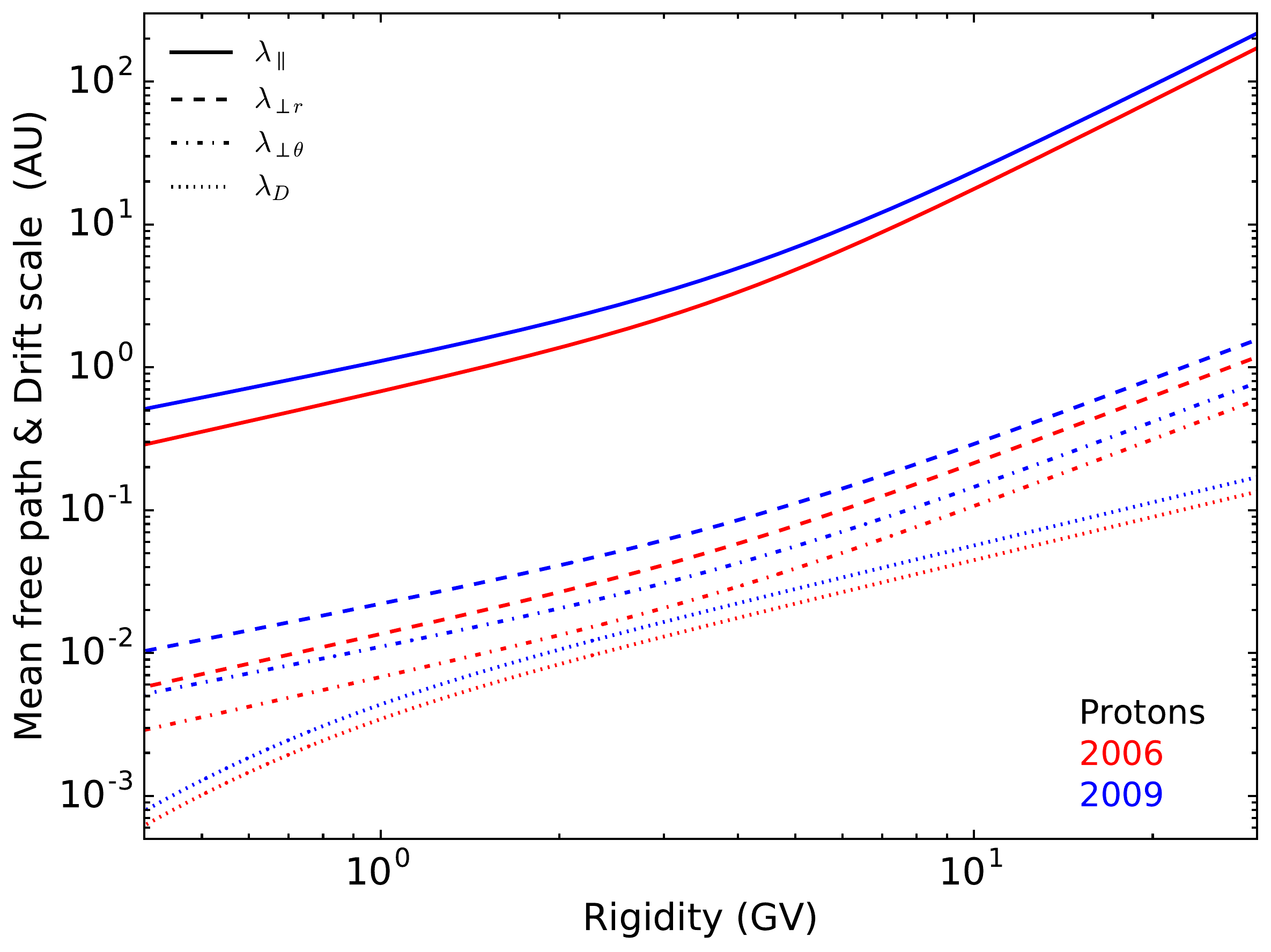}
  \includegraphics[width=0.473\textwidth]{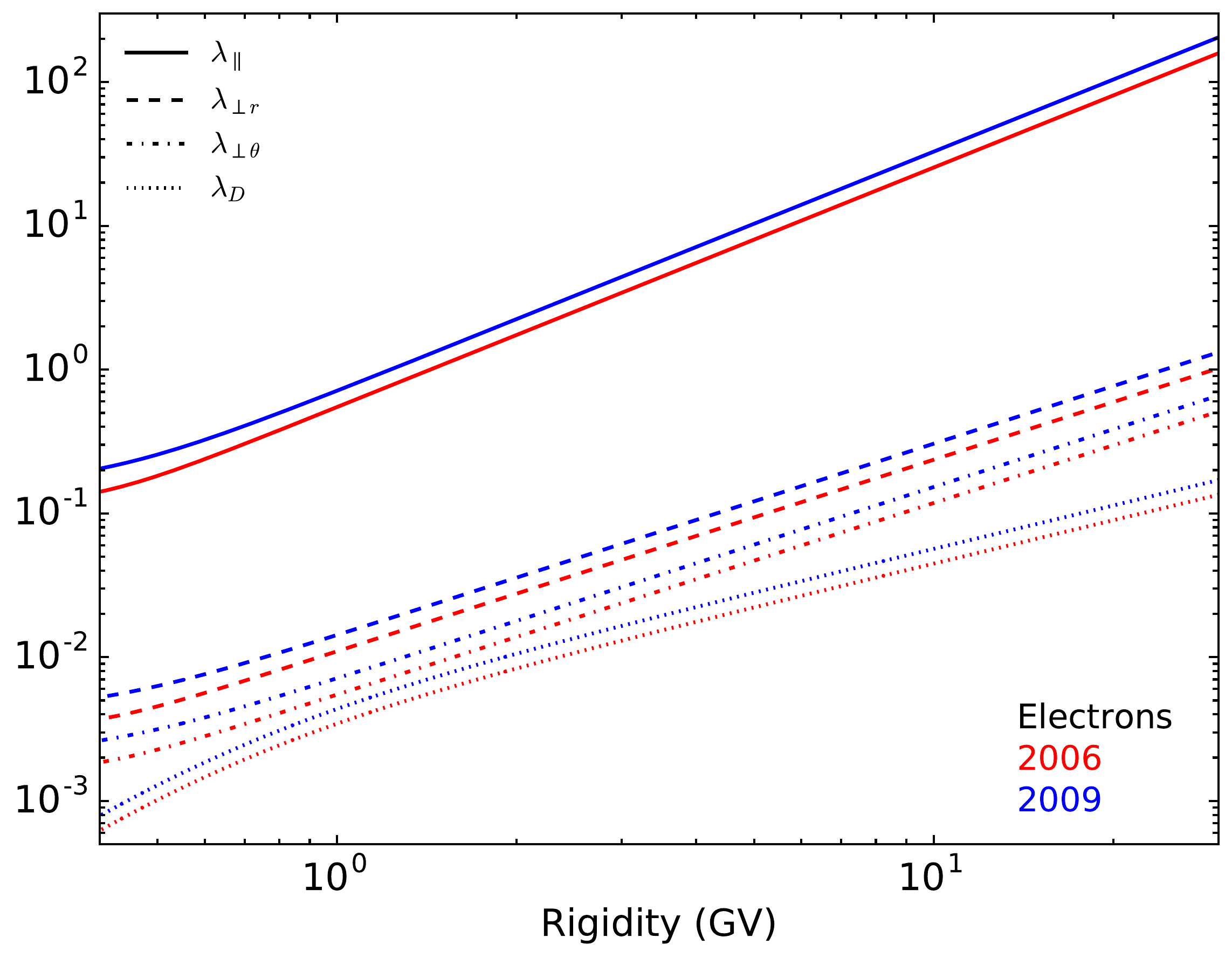}
    \caption{Left panel: the mean free paths (MFPs) and drift scales for the diffusion and drift coefficients, as used in modelling the proton modulation, are shown for the second semesters of 2006 (red) and 2009 (blue).  Parallel MFPs are given by the solid lines, while perpendicular diffusion in the radial and polar directions are given by the dashed and dashed-dotted lines.  The drift scale is shown by the dotted lines. \\ Right panel: Similar to left panel, but for electrons.}  
\label{fig:prot_el_dc}
\end{figure}

Figure~\ref{fig:prot_el_dc} shows the rigidity dependence of the MFPs and drift scales for the second semesters of each year of protons and electrons, for 2006 (red) and 2009 (blue), as obtained from reproducing PAMELA spectra above $0.4$~GV.  Parallel MFPs ($\lambda_\parallel$) are shown by the solid lines, while perpendicular MFPs in the radial ($\lambda_{\perp{r}}$) and polar ($\lambda_{\perp\theta}$) directions are given by the dashed and dashed-dotted lines, respectively.  The drift scales are given by the dotted lines.  From 2006 to 2009, proton and electron MFPs increased by almost the same values above about $2$~GeV as solar modulation conditions became more quiet. Below $2$~GeV this time-dependent increase is
 somewhat larger for protons than electrons. 
As required by turbulence theory \citep[e.g.][]{Burgeretal2000}, proton MFPs have a stronger rigidity dependence above $\sim4\,$GV than below this value.
 The electron MFPs show this behaviour only below about $0.45$~GV. 
Below this rigidity, the electron MFPs become independent of rigidity as discussed by \citet{Potgieter_2015_elmodul}. 
For this study, this feature is less relevant although important for the total modulation of electrons, as is evident in Figure~\ref{fig:H_el_fluxes}.
See also \citet{PotgieterNndanganeni2013}.

\par
\begin{figure}
  \includegraphics[width=0.99\textwidth,height=0.33\textheight]{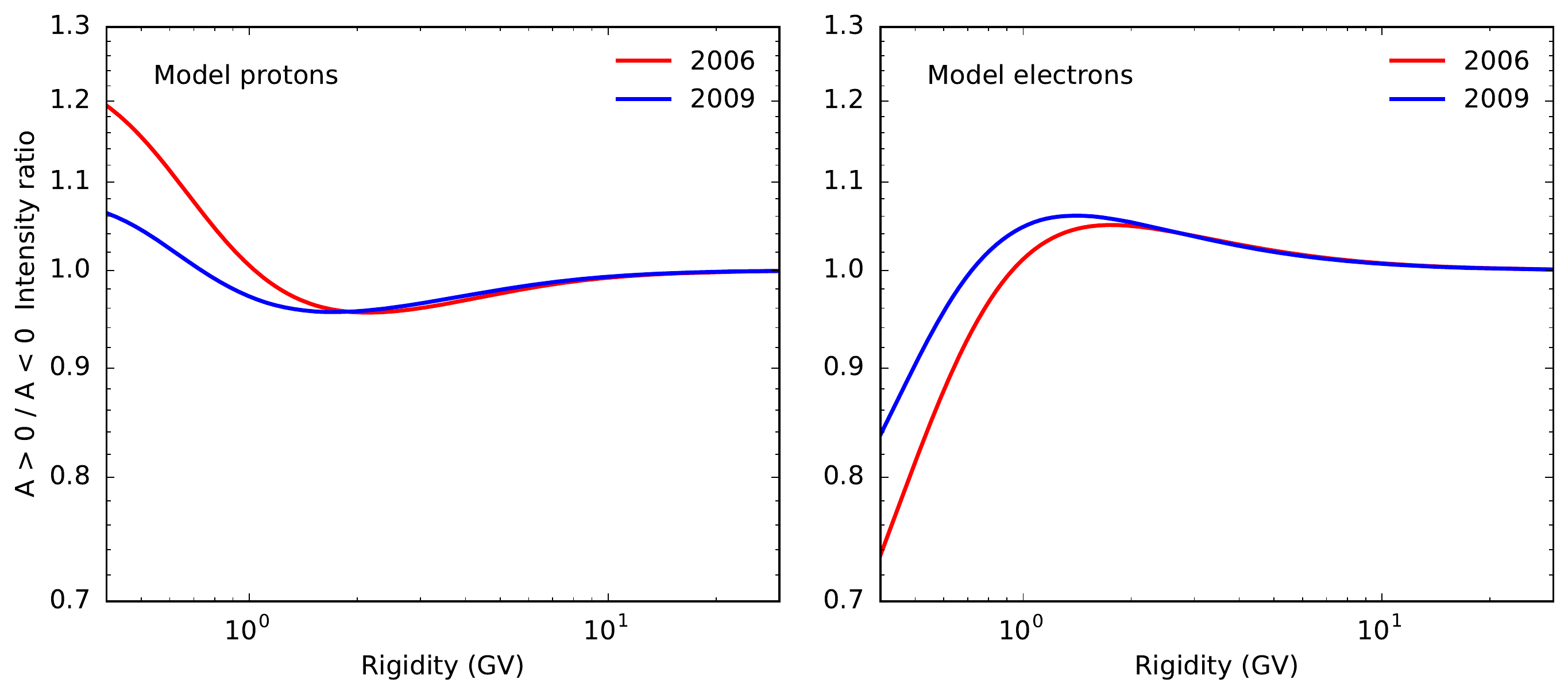}
\caption{The computed ratios of A~$>0$ and A~$<0$ proton (left panel) and electron (right panel) spectra at the Earth (1 AU), for the second semester of 2006 (red lines) and 2009 (blue lines). These ratios are indicative of how large drift effects are in terms of rigidity.}  
\label{fig:theory_1}
\end{figure}

Using a similar numerical model, \citet{Nndanganeni_Potgieter_2016} 
 calculated the effect of drifts on the electron propagation by taking the ratio of $A>0$ spectra to $A<0$ spectra, for the same solar activity conditions.  
Figure~\ref{fig:theory_1} shows similar ratios for proton (left panel) and 
electron (right panel) spectra calculated in this work for the second semester of 2006 (red lines) and 2009 (blue lines).
This is computed by assuming that the same quiet modulation conditions will occur during 
the next solar minimum (A $>$ 0 cycle) than during the solar minimum ($A<0$ cycle) of 2006 to 2009.
It follows from these figures that drift effects for electrons are already significant above $0.4$~GV, subsiding gradually to become less significant above a few GV; 
for protons this is also the case, but notice that drifts cause protons intensities to be higher in the A $>$ 0 cycle than in the A $<$ 0 cycle
 for most of the considered rigidities.  For electrons, on the other hand, the A $>$ 0 intensities are less than for the A $<$ 0 cycle. 
These results thus illustrate the extent of drift effects when solar minimum conditions are present as was the case from 2006 to 2009.
It follows that drift effects become significant from a few tens of MV to a few GV already.
Experimental validations of this numerical model are essential both to determine the rigidity dependence of the modulation parameters and to explore the rigidity range of drifts inside the heliosphere.  
\par
The experimental evidence that we present here consists of showing  how the proton intensities evolved with time compared to the electron intensities
 for 2006 to 2009 and how the differences subside with increasing rigidity related to what the model indicates. 

\section{Discussion}
\label{Sec:discussion}

As mentioned in Section~\ref{Sec:intro}, prior to PAMELA observations there were experimental evidence of charge-sign dependent solar 
modulation effects being present.  In particular, the Ulysses mission provided the opportunity to study the long-term propagation and modulation of galactic CRs
through measurements of electrons (sum of \el and \pos), protons and helium in specific energy channels below a few GeV~\citep{Heber_2002, Heber_2009}. 
This mission, launched on 6 October 1990, followed a highly inclined (80.2 degrees) elliptical orbit around the Sun until the switch-off in June 2009, 
being the only spacecraft exploring high-latitude regions of the inner heliosphere.
The intensity variation observed along its orbit was reproduced by the modulation modelling from~\citet{Ferreira_2004}, including all the main physical transport
 processes in the heliosphere. The measured variation of the $\mbox{${\rm e}/{\rm p}\;$}$ ratio, especially its latitudinal variation, was interpreted in terms of both temporal and spatial effects, 
indicating that drift effects are important over a large part of the 11-year solar cycle. \citet{Ferreira_2004} also emphasized the importance
 of simultaneous measurements of protons and electrons in order to understand and appreciate drift effects.
See also the review by \citet{HeberPotgieter2006}.

\begin{figure}
\epsscale{1.1}
\plotone{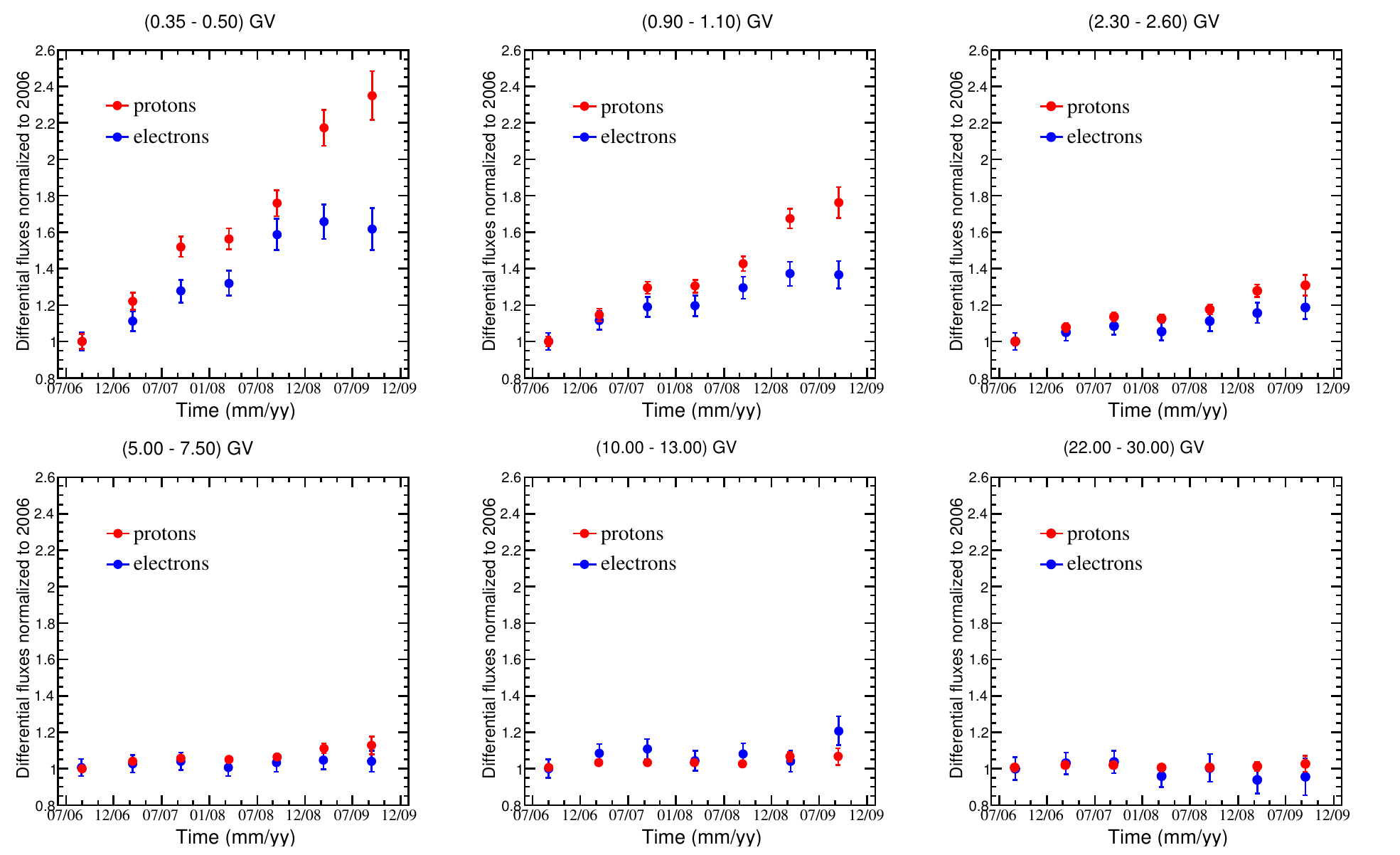}
\caption{Normalized electron and proton intensities, measured from July 2006 to December 2009 for a selection of six rigidity intervals, increasing from left to right, top to bottom as indicated.
The intensity-time profiles are normalized with respect to the averaged intensity measured from July to November 2006.}
\label{fig:6_rigidities}
\end{figure}

The PAMELA low-energy data 
discussed in Section~\ref{Sec:data} 
extend the study of the temporal evolution of electron (${\rm e^{-}}$) and proton intensities over a much wider rigidity range than before.
 Figure~\ref{fig:6_rigidities}  illustrates how the proton and electron intensities changed with time from the second half of 2006 to the end
of 2009, for several selected rigidity intervals. In order to compare the fluxes at the same rigidities, we use the electron rigidity as a reference
and select the corresponding proton intensity by performing a linear interpolation of the measured proton spectrum.
Each panel refers to a different rigidity interval, increasing from top to bottom, left to right.
Intensities are normalized to the values measured between July 2006 and November 2006. 
The error  bars are the quadratic sum of the statistical and systematic uncertainties. 
The results shown in this figure emphasize that electron and proton intensities developed differently over the last solar minimum period, 
showing a significant steeper recovery trend towards solar minimum modulation for protons than for electrons in response to solar modulation conditions.
Additionally, it shows how these differences dissipate with increasing rigidity.
The top-left panel presents the measurement at the lowest rigidity available for comparison, between $0.35$ and $0.5$~GV. 
In this rigidity range protons increased by a factor $\sim 2.4$ over $\sim3$ years, while electrons increased by only a factor $\sim1.6$ over the same period.

\begin{figure}
\epsscale{1.}
\plotone{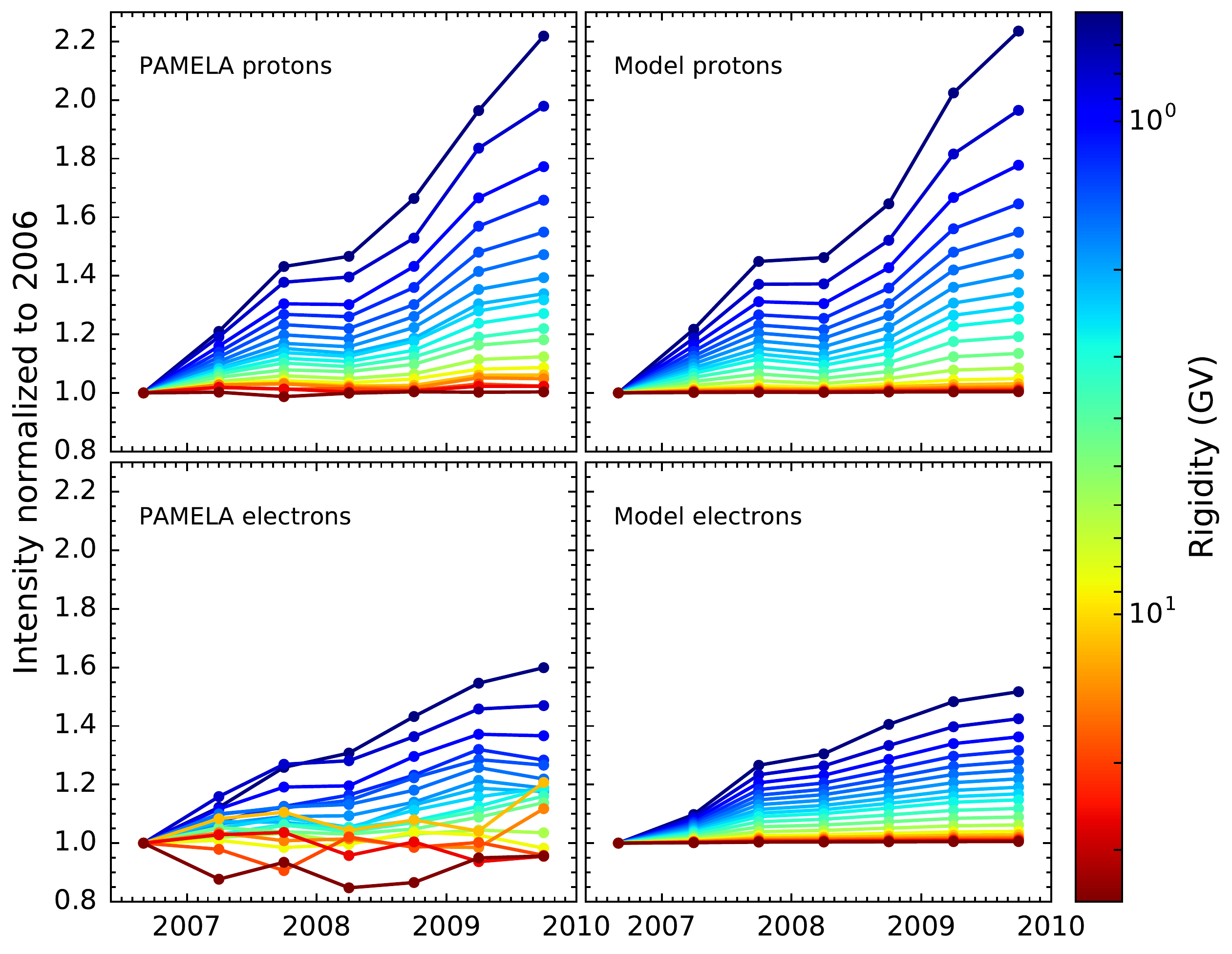} 
\caption{Comparison of differential intensity measurements from PAMELA (left panels), and computed intensities from the model (right panels), for protons (top panels) and electrons (bottom panels).  Intensities are shown as a function of time, normalized to the second semester of 2006.  Rigidity values are indicated by the different colours, according to the colourbar, with high rigidities (around $30\,$GV) represented by the red coloured lines, and low rigidities (around $1\,$GV) represented by the blue coloured lines.}  
\label{fig:prot_elec_time}
\end{figure}

Comparing the data with the model results shows that the model is able to reproduce the proton and electron intensity increase observed by PAMELA
 from 2006 to 2009, in particular the larger relative increase for protons compared to electrons, which is indicative of drifts.
Figure~\ref{fig:prot_elec_time} compares PAMELA proton (upper left) and electron (lower left) measurements with computed proton (upper right) and electron (lower right) intensities from the model, at the same rigidities. Similarly to Figure~\ref{fig:6_rigidities}, PAMELA and model intensities are shown as a function of time, normalized to the last semester of 2006.
It's worth to note that only when including drift effects in a comprehensive 3D modulation model, these charge-sign modulation effects can be reproduced, 
since differences in the other modulation coefficients are not sufficient to account for the observed behaviour.

The differences reported in Figures~\ref{fig:6_rigidities} and \ref{fig:prot_elec_time} are interpreted as an indication of particle drifts: the protons mainly drift in along the equatorial regions of the heliosphere, encountering and following the wavy HCS. When the waviness slowly subsides with less solar activity, the tilt angle follows and is reduced, making the modulation conditions more favourable for protons to reach the Earth so that their intensity increases faster than the electron intensity which mostly drift inwards through the polar regions of the heliosphere thus mostly escaping the changes in the waviness of the HCS. During this time, the electrons reach their maximum intensity levels sooner than the protons, and level off already by the end of 2008 whereas the proton intensity keeps increasing as the tilt angle keeps decreasing. 
\par
Figure~\ref{fig:6_rigidities} shows that with increasing rigidity (from top to bottom, left to right), this difference in the behaviour of protons and electrons gradually diminishes to become less pronounced 
and eventually seems to subside completely above $\sim11$~GV.
This observed phenomenon is interpreted as an indication of how drift effects subside with increasing rigidity while solar activity decreases. 
This seems to be in contradiction with Eq.~\ref{eq:eq5}, according to which the drift coefficient scales proportional to rigidity at larger rigidities.
However, whereas the drift coefficient, and the corresponding CR drift velocity in Eq.~\ref{eq:eq4} increase, the other modulation processes reduce the global intensity gradients between the LIS value and the intensity at the Earth with decreasing solar activity, so that drift effects effectively diminish. This subtle modulation effect is described in detail by~\citet{Nndanganeni_Potgieter_2016}. 
PAMELA observations indicate that during the solar minimum period from 2006 to 2009, drift effects where not observable beyond about 10-13 GV.

\begin{figure}
\epsscale{1.}
\plotone{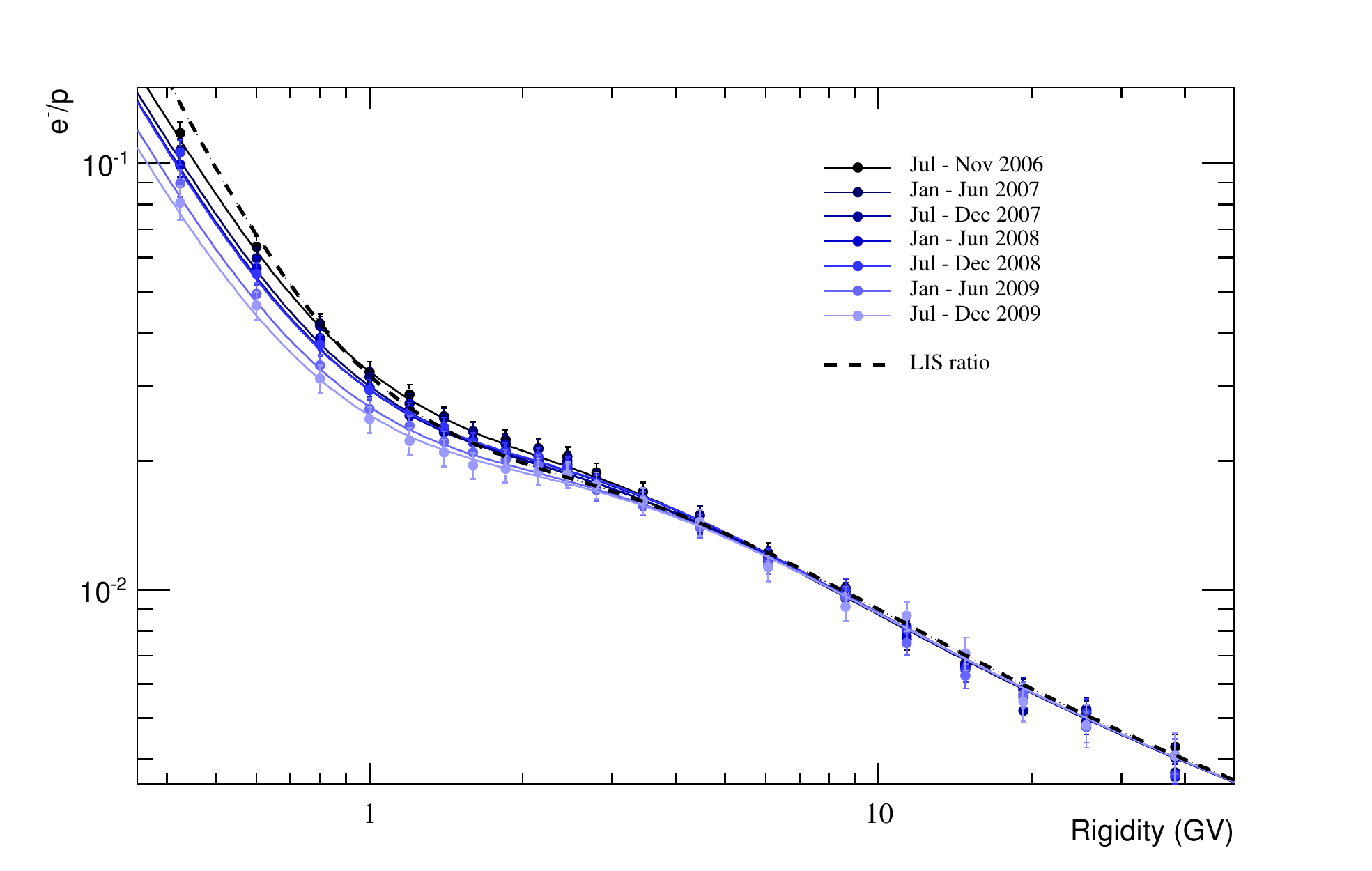}
\caption{Electron to proton ratio measured by PAMELA from 2006 to 2009 indicated by data points with error bars. 
Solid lines represent the modelling results for the same periods, while the dashed line is the electron to proton LIS ratio.}  
\label{fig:fluxratio}
\end{figure}

 Figure~\ref{fig:fluxratio} shows the rigidity dependence of the \ep ratios for the spectra measured by PAMELA (symbols) and computed spectra (solid lines) for each semester, from July 2006 to December 2009, illustrating the observed differences between the two oppositely charged CRs.
At lower rigidities a decreasing trend in the \ep ratio is observable as the heliospheric conditions change.
The dashed line represents the \ep ratio
based on the respective LIS. The ratios given by the computed spectra give a good representation of the observed PAMELA \ep ratios with regard to shape and values over 
the studied rigidity range. Evidently, as solar modulation gets less with increasing rigidity, the \ep ratio slope approaches the LIS ratio. 
These ratios show that charge-sign modulation is
largest at the lowest rigidity interval observed ($350$~-~$500$~MV) in agreement with the model. However, the models give that drift effects should reach a maximum around $100$~MV
 for both electrons and protons~\citep{Nndanganeni_Potgieter_2016}.

\section{Summary and Conclusions}

Simultaneous observations of oppositely charged cosmic rays have been made available by the PAMELA space spectrometer, that measured particle spectra during the entire A $<$ 0 polarity epoch minimum of solar cycle 23/24. We compared proton and electron spectra measured by PAMELA for rigidities between 400 MV and 50 GV. Data show that, as solar activity gradually decreased from 2006 and 2009, low energy protons presented a different behavior compared to electrons over time, with proton intensities responding more to changing heliospheric conditions at the same rigidities. We studied the spectra evolution for both particles with a comprehensive 3D modulation model, which accounted for all the important modulation processes, including drift. We conclude that the observed effect, subsiding with increasing rigidity, is mostly caused by drifts, providing new and additional evidence of charge-sign dependent solar modulation during a very quiet solar activity condition period.

\acknowledgments

Acknowledgments

M. S. P. and E. E. V. acknowledge partial
financial support from the the South African National
Research Foundation (NRF) under their Research
Cooperation Programme. 
E. E. V. thanks the Space Science Division of the South African Space Agency (SANSA) 
for partial financial support during his PhD-studies.
R. M. acknowledges partial financial support from The Italian Space Agency (ASI) under 
the program "Programma PAMELA - attivita' scientifica di analisi dati in fase E".
The authors would like to thank Dr. M. Boezio for the useful discussion and helpful suggestions.
Part of this work is based on archivial data provided by the ASI Science Data Center (ASDC). 
\clearpage

\bibliography{bibliography}{}
\bibliographystyle{aasjournal}



\end{document}